\documentclass{PoS}

\title{Towards enhanced databases for High Energy Physics}

\ShortTitle{Towards enhanced databases for HEP}

\author{Andrea Ceccarelli\\
  Dipartimento di Matematica e Informatica,
  Universit\`a degli Studi di Firenze,\\
        E-mail: \email{andrea.ceccarelli@unifi.it}}

\author{Andrea Cioni\\
  Dipartimento di Matematica e Informatica,
  Universit\`a degli Studi di Firenze,\\
          E-mail: \email{andrea.cioni@stud.unifi.it}}

\author{\speaker{Maria Vittoria Garzelli}\\
Dipartimento di Fisica e Astronomia,
  Universit\`a degli Studi di Firenze,  \& INFN, Firenze, \\
  Institut f\"ur Theoretische Physik,
  Eberhard Karls Universit\"at T\"ubingen, \\
        E-mail: \email{maria.vittoria.garzelli@cern.ch}}

\author{Piergiulio Lenzi\\
Dipartimento di Fisica e Astronomia,  Universit\`a degli Studi di Firenze,  
\& INFN, Firenze,\\
          E-mail: \email{lenzip@fi.infn.it}}

\author{Laura Redapi\\
  GARR Consortium,\\
Dipartimento di Fisica e Astronomia,  Universit\`a degli Studi di Firenze, \\
          E-mail: \email{laura.redapi@gmail.com}}

\abstract{
The accumulation of a large amount of new experimental data at an impressive rate at present and future collider experiments has led to important questions concerning data storage and organization, their public access and usability, as well as their 
efficient usage in order to discriminate between different theories. For the last fourty years, the HEPData database has been the reference database for the worldwide community of elementary particle physicists, from DIS to fixed-target and collider experts. Using as a basis a dump of HEPData\footnote{the HEPdata dump on which this work is based was kindly provided us by K. Ellis, F. Krauss and G. Watt (IPPP, Durham), with whom further collaboration is foreseen for the future.}, we discuss possible paths to enhance the capabilities of databases for High Energy Physics. Our starting point is the reorganization of the data in a different scheme, which allows for the application of OLAP techniques to automatically extract information at a multidimensional level, answering to complex queries. The feedback of the DIS community is important for understanding specific needs, aiming at a more effective storage, extraction and presentation of the data and information of their interest.}

\FullConference{XXVII International Workshop on Deep-Inelastic Scattering and Related Subjects - DIS2019\\
		8-12 April, 2019\\
		Torino, Italy}

\begin{document}

\section{Introduction}

Physics experiments are accumulating data at an impressive rate, and questions about the most efficient ways of storing, accessing and preserving these data in view of their future re-usage and of the possibility of performing consistent comparisons of the results of different experiments, are open. In this respect, the Group of scientists involved in the discussion of Data Preservation in High Energy Physics (DPHEP), considering the flow of information in a typical experimental analysis at an high-energy accelerator, has suggested a classification of data into four different levels, characterized by different contents~\cite{Akopov:2012bm}. In particular, going from larger data sizes to lower ones, they distinguish raw data collected during the measurement and reconstruction software (level~4), analysis-level data (level 3), i.e. the data and software sufficient to perform a complete re-analysis, simplifed-event-level data (level 2), i.e. sets of vectors describing the particles of each selected event in a simplified format, and published data (level 1), i.e. the data reported in scientific works. 

In this work, we focus on level-1 data, which are characterized by the fact of being 1) multisource, corresponding to the outcome of many different experiments, 2) multidimensional and 3) heterogeneous. This means that they are characterized by different dimensions, whose number vary according to the observable/experiment considered and the way of reporting the information, and they are organized in different kinds of data structures. These features represent in general a challenge for the organization of all these data in a common database and their retrieval, also considering that information of the same quality might have been reported by different experiments in different ways (for example, the information concerning correlated uncertainties, crucial for QCD analyses interesting for the DIS community, can be stored in covariant matrices or in other formats) and that information reported in some analyses can be absent in other ones.
 
The most complete open-access collection of level-1 data on high-energy scattering reactions at experiments conducted from the early days of particle physics to present, is the HEPData database, which can be regarded as the reference database for the worldwide community of elementary particle physicists at accelerators. This database, mostly developed at the Institute for Particle Physics Phenomenology (IPPP) of the University of Durham and, more recently, also in collaboration with the Open Access group at CERN, has a long history, from the sixties to the present. A legacy version~\cite{Stirling:1993gc} was organized according to the hierarchical Berkeley Database Management System (DBMS), with information stored in tree-like structures, with papers (including experimental data) as record units. The user could access the stored information through Fortran routines. This version was superseded by a more modern one, called HepData~\cite{Buckley:2010jn}, developed within the CEDAR project~\cite{Butterworth:2004mu}, with the information reorganized according to the relational DBMS MySQL, in a complex structure of many tables with different kinds of relations (one-to-many, many-to-many). At that time, 
the HepML data format~\cite{Belov:2010xm}, based on XML, was created to enter data, as well as a web interface in Java for allowing user queries~\cite{HepDataweb}. The parallel development of further tools, like e.g. Rivet and JetWeb in the CEDAR context~\cite{Buckley:2006np}, was driven by the ultimate objective of automatizing the comparison of the collected data with theory predictions as much as possible. Finally, the most recent HEPData version~\cite{Maguire:2017ypu} has been developed in the last few years. Differently from previous versions, hosted on Durham servers, the last version~\cite{HEPDataweb} is hosted on the CERN OpenStack infrastructure. It makes use of the PostgreSQL DBMS and of the Invenio3 open-source framework for large-scale digital repositories. An interface has been created to allow the experimental collaborations to fill the database with their data, in files written in {\texttt{yaml}}, a superset of the {\texttt{json}} markup language. The current emphasis is on the inclusion of more data, not only limited to HEP reactions, and on the development of statistical models and of studies of Physics Beyond the Standard Model.     

 Our work starts from the analysis of the present status of HEPData, and shows how a re-organization of the data into a different structure, inspired by OnLine Analytical Processing (OLAP) techniques~\cite{kimball:2013}, allows the user to perform queries, centered on the extraction and comparison of data from different papers, more complex than those allowed by the present HEPData structure.  

\section{The MineHEP search-engine prototype}

Our final goal is organizing the already available level-1 HEP data in a way
to automatically extract from them as much information as possible. 
Using as a starting point the MySQL dump file \texttt{hepdatapublic.dmp.gz\_01Apr18} of HEPData, we have built a search-engine prototype, dubbed MineHEP, capable of queries more complex than those possible nowadays through the HEPData web interface. 

 In relational databases, data are organized in tables structured in fields, where each field corresponds to a column and each data entry to a row. There is not a unique recipe to organize the tables and to connect them among each others, i.e. to define the structure of a database. However, the information retrieval capabilities and performances are strictly connected to the structure of the database. 

\begin{figure}
\begin{center}
\includegraphics[width=0.95\textwidth]{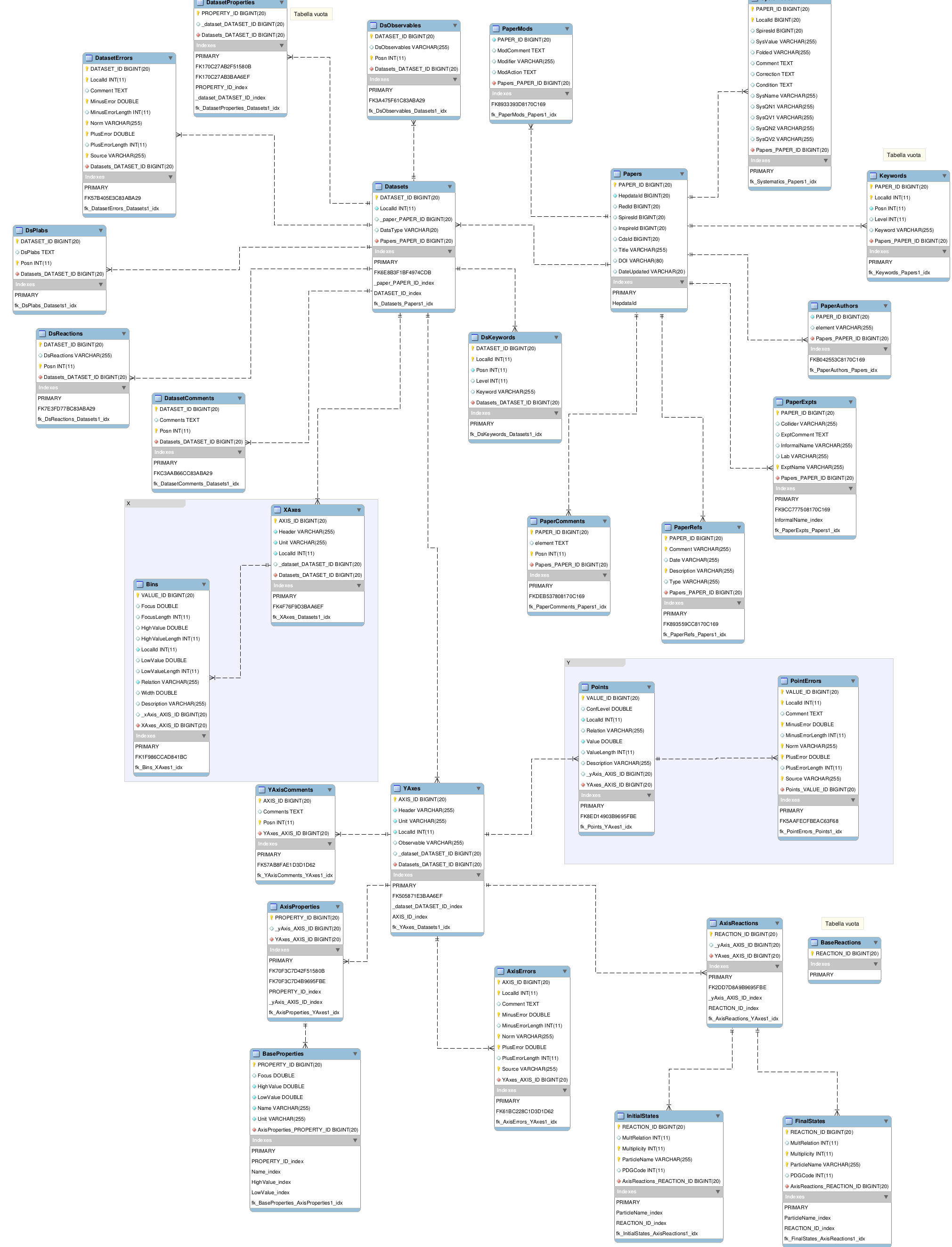}
\caption{\label{old} Structure of the HEPData database, extracted from the dump file dating back to April 2018, including all data up to October 2017.}
\end{center}
\end{figure}

The structure of HEPData, as inferred from the aforementioned dump, is shown in Fig.~\ref{old}, where one can see that there are two prominent tables called {\it Datasets} and {\it Papers}, connected between each other. Each of them is then connected to many other tables, which, in turn, can be connected to further ones. Although this structure and the relations between tables reflect an intuitive way of organizing the information provided by the experimental collaborations in papers including their data, the complexity of the schema makes it difficult, if not impossible, to write and efficiently execute queries capable of extracting information from tables distant many hops from each other.     

\begin{figure}
\begin{center}
\includegraphics[width=0.69\textwidth]{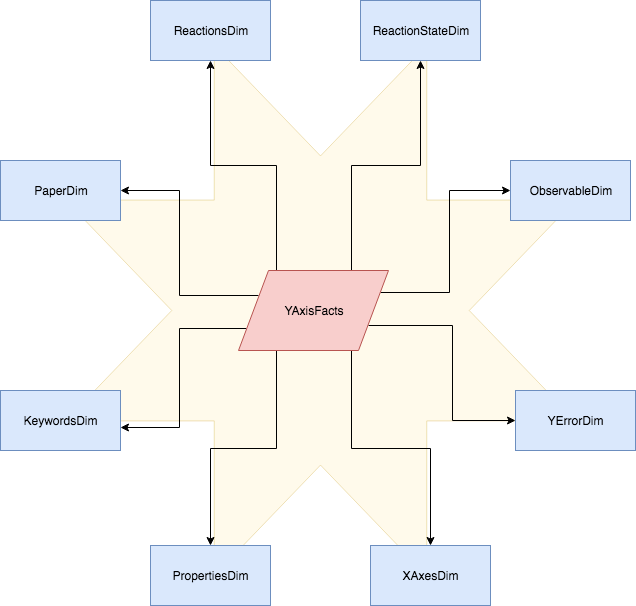}
\caption{\label{star} Star-schema structure at the basis of the 
MineHEP search enegine.}
\end{center}
\end{figure}

In order to allow for complex queries, we have reorganized the information 
in a star-schema, as shown in Fig.~\ref{star}.
Considering each data point as a collection of (\{$x_i$\}, $y$) values, where $y$ is a function of the \{$x_i$\} coordinates, we have chosen to center the star-schema around the table of the $y$-values of all data points published by the experiments, 
to ensure the maximum granularity in information retrieval.  The central table, also called {\it fact} table according to the OLAP terminology, is surrounded by {\it dimensional} tables, each of which refers to a group of attributes associated to each entry of the fact table. We found it convenient to reorganize the relevant information concerning these attributes in eight dimensional tables, called 
1) {\it XAxesDim}, containing the information on the \{$x_i$\} coordinates, to which each value $y$ stored in the fact table refers; 
2) {\it YErrorDim}, containing the information on the uncertainties on $y$; 
3) {\it ObservableDim}, containing the information on the observable to which the $y$ value refers;
5) {\it ReactionDim}, containing the information on the reaction in which the datum was measured; 
4) {\it ReactionStateDim}, containing the information on each single particle in the initial and final state of this reaction;  
6) {\it PropertiesDim}, containing the information on the properties of the reaction, from the center-of-mass energy to the analysis cuts;
7) {\it PaperDim}, containing various information (arXiv number, doi, collaboration, etc.) on the paper to which the data point belongs; 
8) {\it KeywordDim}, containing the keywords associated to each data point. 

\begin{figure}
\begin{center}
\includegraphics[width=1.0\textwidth]{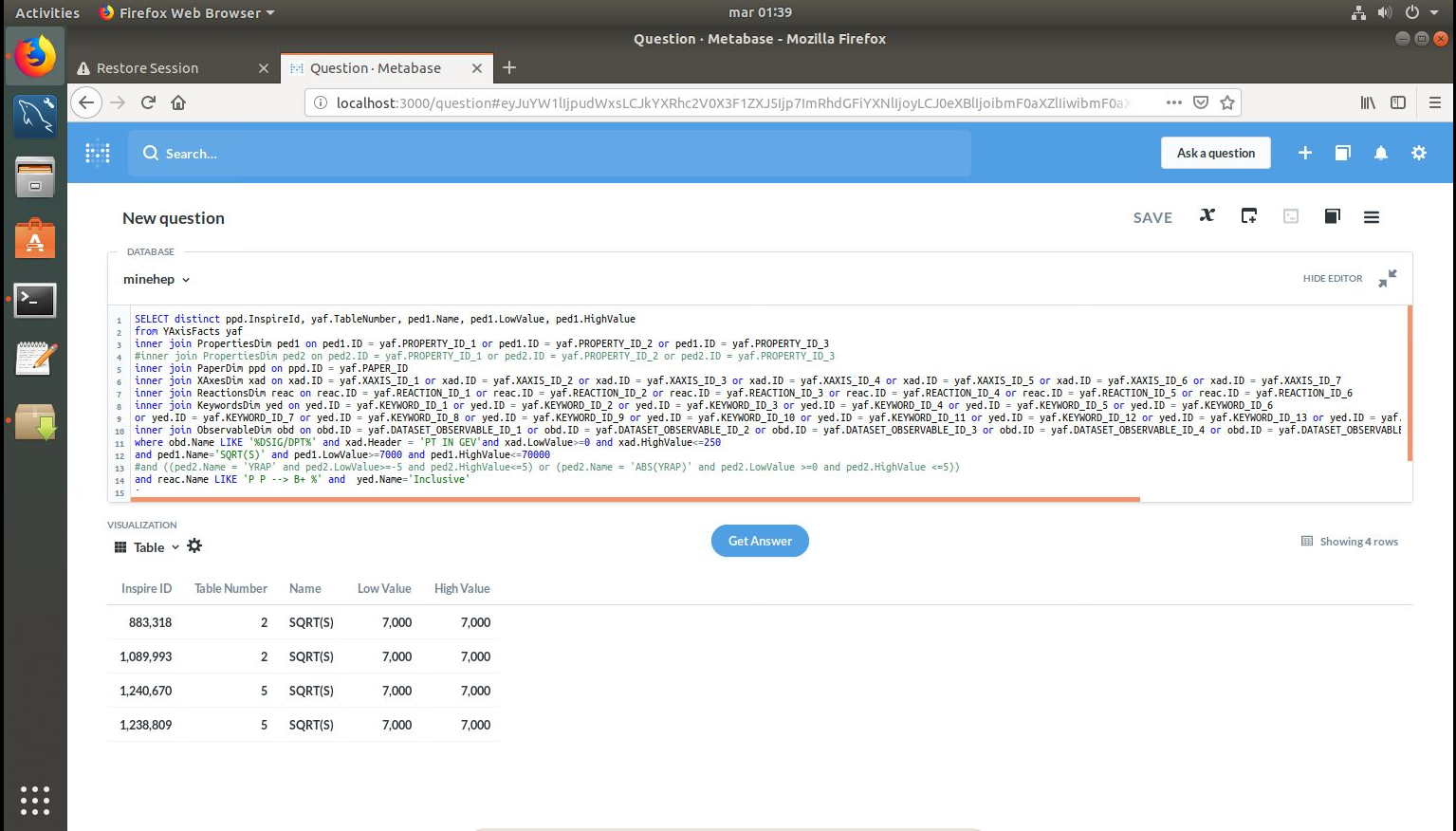}
\end{center}
\caption{\label{querymv} Example of a MySQL query to the MineHEP search engine, executed through its preliminary interface to Metabase~\cite{metaba}. It
selects
 all tables of papers reporting the $p_{T,B^+}$ distribution in inclusive $B^+$ meson production in $pp$ collisions at center-of-mass energies $\sqrt{s}$ = 7 TeV, for 0 $<$ $p_{T,B}$ $<$ 250 GeV. The output of the query, reporting the InspireID of the paper and the relevant Table number, is also shown.}
\end{figure}

We have started testing the reliability and the robustness of the implementation, by developing a set of queries of increasing complexity, which can be interesting for typical HEP users. An example of a query for selecting data on inclusive production of $B^+$ mesons in $pp$ collisions, interesting for DIS experts aiming at performing PDF fits, is reported in Fig.~\ref{querymv}. The complexity of this query can be further increased by requiring that the experimental points are obtained under additional analysis selection cuts. 

In case of queries involving multiple entries of a same dimensional table (e.g. multiple keywords, or multiple properties), it is worth exploring the performance of variants of the star schema, making use of intermediate bridge tables. We are currently working on this topic.    
  
In the future we plan to host the MineHEP search engine prototype on a public server, granting access to users, who will be able to profit of its query capabilities and will provide very useful feedback for its further development. 

{\bf Acknowledgements} We are grateful to the members of IPPP, Durham, for useful discussions at different stages of this project.

\end{document}